\RequirePackage{amsthm}
\documentclass[referee,sn-mathphys,Numbered]{sn-jnl}

\usepackage{float}
\usepackage{lscape}
\usepackage{caption}
\usepackage{subcaption}

\usepackage{graphicx}%
\usepackage{multirow}%
\usepackage{amsmath,amssymb,amsfonts}%
\usepackage{amsthm}%
\usepackage{mathrsfs}%
\usepackage[title]{appendix}%
\usepackage{xcolor}%
\usepackage{textcomp}%
\usepackage{manyfoot}%
\usepackage{booktabs}%
\usepackage{algorithm}%
\usepackage{algorithmicx}%
\usepackage{algpseudocode}%
\usepackage{listings}%

\theoremstyle{thmstyleone}%
\theoremstyle{thmstyletwo}%

\theoremstyle{thmstylethree}%

\begin{document}

\title[Electromagnetic flux in wormholes]{Outgoing electromagnetic flux from rotating wormholes}


\author*[1]{\fnm{Milos} \sur{Ertola Urtubey}}\email{meusay@fcaglp.unlp.edu.ar}

\author[1]{\fnm{Daniela} \sur{P\'erez}}

\author[1,2]{\fnm{Gustavo E.} \sur{Romero}}


\affil*[1]{\orgname{Instituto Argentino de Radioastronom\'ia (IAR, CONICET/CIC/UNLP)},  \city{Villa Elisa}, \postcode{C.C.5, (1894)}, \state{Buenos Aires}, \country{Argentina}}

\affil[2]{\orgdiv{Facultad de Ciencias Astron\'omicas y Geof\'sicas}, \orgname{Universidad Nacional de La Plata}, \city{La Plata}, \postcode{1900}, \state{Buenos Aires}, \country{Argentina}}



\abstract{We show for the first time that rotating wormholes are capable of emitting a Poynting flux in the process of accreting magnetized matter. To this end, we analyze the Damour-Solodukhin metric describing a Kerr-type wormhole and calculate the electromagnetic flux assuming a specific geometry for the magnetic field contained by the wormhole ergosphere. We find that for highly rotating wormholes a mechanism similar to that of Blandford and Znajek is possible, and the emitted electromagnetic flux is of the same order as for a Kerr black hole.}




\maketitle

\section{Introduction}

Many would agree that general relativity (GR) is a theory that stands out for its amazing predictive power. Black holes and gravitational waves, for example, were discovered theoretically decades before their existence was confirmed. However, there are several other predictions derived from Einstein's field equations that are still in the realm of theory: wormholes are one of the best examples.

Wormholes are regions of spacetime with non-trivial topology\footnote{The word ``topology'' is sometimes used loosely in the literature. Informally, topology is the branch of mathematics that deals with the properties of a geometric object that are preserved under continuous deformations such as stretching, twisting, crumpling, and bending; that is, without closing holes, opening holes, tearing, sticking, or passing through itself. More formally, ${\cal T}$ is a topology on a set $M$ if 1) ${\cal T}$ is a collection of open subsets of $M$, 2) $\emptyset \in {\cal T}$ and $M\in {\cal T}$, 3) any arbitrary (finite or infinite) union of members of ${\cal T}$ belongs to ${\cal T}$, and 4) the intersection of any finite number of members of ${\cal T}$ belongs to ${\cal T}$. A sphere and a piramid have the same topology because one can be continuously deformed into the other. Of course, they have different geometries. Geometry deals with quantitative aspects, which can be quantified by assigning a metric to the topological space. It is important to note, however, that the metric structure alone does not allow us to determine the topology of a set. This is why we cannot find the topology of the universe by solving Einstein's equations. }; they are shortcuts between different regions of the universe. Unlike black holes, wormholes are characterized by the absence of event horizons, so they form "bridges" connecting different events in spacetime. A wormhole has a minimum radius, the throat; in their simpler versions, two mouths on either side of the throat allow the passage of matter and fields in both directions.

Since Morris and Thorne's seminal paper on transversable wormholes \citep{mor+88}, there has been intense theoretical research on the subject. There is a wide variety of solutions to wormholes, both in GR and in alternative theories of gravity \citep{vis95,2017Lobo}. There are also cosmological wormhole solutions, that is, wormholes that are coupled to the cosmic background dynamics\footnote{Note that all these cosmological metrics are spherically symmetric. None of them, including the solution of Perez and Raia Neto \citep{per+23} which generalizes the old Kim \citep{kim96} solution, has an ergoregion. As we will discuss next, the presence of an ergosphere is an essential feature for the generation of a Poynting flux by the Blandford-Znajek mechanism. There may be another type of astrophysical manifestations associated with cosmological wormholes that, to our knowledge, have not yet been explored. } \citep{kim92,rom93,kim96, bah+16, per+23}. 

There is also a line of research devoted to the possible astrophysical manifestations of wormholes, such as gravitational lensing \citep{Torres1998PhRvD,Safonova2001PhRvD,ovg+19} (a review on astrophysical wormholes can be found in \citep{bam+21}). An almost unexplored consequence of a wormhole immersed in a region of matter and fields is the possibility of producing astrophysical jets\footnote{While preparing this manuscript, it was published the first dynamical model of plasma accretion onto traversable wormholes by performing 3D General Relativistic Magneto-Hydrodynamical (GRMHD) simulations of the flow on both sides of a wormhole \citep{com+24}. They showed that part of the magnetized plasma that accumulates in the throat is launched in as a mildly relativistic thermal wind. It should be noticed that they consider the Simpson-Visser metric that represents a spherically-symmetric non-rotating wormhole, so the Blandford-Znajek mechanism does not take place (there is no ergoregion). However, their GRMHD simulations point out that wormholes can launch particles and fields, even in absence of rotation. }. Jets seem to be associated with accretion onto a rotating compact object. Another basic ingredient, at least for relativistic jets, is the presence of large-scale magnetic fields. Currently, there are several models to explain the launch, collimation, and acceleration of jets. One of the most studied is the Blandford-Znajek mechanism \citep{1977Blandford}, which requires the presence of a central rotating object and a magnetosphere. In this
mechanism, the ergosphere, and not the event horizon, is the fundamental component, as shown by analytical results and simulations \citep{2001Komissarov, 2002Komissarov, 2004Komissarov}. Therefore, spacetimes in which an ergospheric region is present, as in the case of rotating wormholes, could produce a net Poynting flux, and perhaps relativistic jets.

To our knowledge, there are only a few spacetime metrics that represent rotating wormholes in GR: the generalization of the Ellis wormhole \citep{2014Kleihaus}, the rotating wormhole solution of Teo \citep{1998Teo} and the Damour-Solodhukin Rotating Wormhole (DSRW) metric \citep{2018Bueno, 2019Karimov}. There are also solutions where matter with angular momentum sets the throat of an Ellis wormhole in rotation \citep{hoff+18,hoff+19}, rotating wormholes in AdS \citep{cas+18} and an overcharged Kerr-Newman-NUT (KNN) wormhole solution \citep{cle+23}. In this work, we will focus on the DSRW metric: it has an ergoregion and it resembles the Kerr metric, a feature that will be useful for comparison with known results. 

The goal of this work is to determine the conditions under which the DSRW can produce a net Poynting flux in the process of accreting matter with magnetic fields. Then, we will compare the obtained results with the analogous ones for a Kerr black hole of equal mass and angular momentum in order to assess whether the production of astrophysical jets is a viable process in wormholes.


\section{Rotating wormhole spacetime}\label{sec:2}

The Kerr-type wormhole metric derived in \cite{2018Bueno} is characterized by three parameters: the wormhole mass $M$, the spin $a\equiv J/M$, and the deformation parameter $\lambda$. In Boyer-Lindquist coordinates $(t,r,\theta, \phi)$ the line element takes the form

\begin{equation}
    \begin{split}
        ds^{2}= & -\left(1-\frac{2GMr}{c^{2}\Sigma}\right)c^{2}dt^{2}-\frac{4GMar\sin{\theta}^{2}}{c\Sigma}dtd\phi+\frac{\Sigma}{\hat{\Delta}}dr^{2} \\
        & +\Sigma d\theta^{2}+\left(r^{2}+a^{2}+\frac{2GMa^{2}r\sin{\theta}^{2}}{c^{2}\Sigma}\right)\sin^{2}{\theta}d\phi^{2}.
        \label{4.1}
    \end{split}
\end{equation}

Here, $\Sigma$ and $\hat{\Delta}$ represent auxiliary functions, $\Sigma \equiv r^{2}+a^{2}\cos^{2}{\theta}$, $\hat{\Delta} \equiv r^{2}-2GM/c^{2}(1+\lambda^{2})r+a^{2}$. The coordinate $r$ is defined on an interval $r_{+}\leq r<\infty$, where $r_{+}$ is the largest solution of the equation $\hat{\Delta}=0$. In the limit $\lambda \to 0$, the Kerr black hole metric is recovered.

The Ricci tensor has 6 non-zero components

\begin{equation}
    R_{\mu\nu}=\lambda^{2}f_{\mu\nu}(r,\theta),\quad \mu=\nu=t,r,\theta,\phi \quad \text{o} \quad \mu=t,\,\nu=\phi,
\end{equation}

being $f_{\mu\nu}(r,\theta)$ a function that only depends on $M$, $a$ and the coordinates $r$ and $\theta$. Since the Kerr metric is a vacuum solution of GR, the parameter $\lambda$ is related to the matter content and thus can be interpreted as a measure of the deviation from the Kerr metric.

The line element \eqref{4.1} has two apparent singularities for those values of $r$ and $\theta$ such that $\Sigma=0$ and $\hat{\Delta}=0$, representing the two boundary surfaces of the wormhole, the ergosphere and the throat. 

In Boyer-Lindquist coordinates, the ergosphere and the throat follow the expressions 

\begin{eqnarray}
    r_{S^{\pm}}&=&\frac{GM}{c^{2}}\pm \sqrt{\left(\frac{GM}{c^{2}}\right)^{2}-a^{2}\cos^{2}{\theta}},\label{2.124}\\
    r_{\pm}&=&\frac{GM}{c^{2}}(1+\lambda^2)\pm \sqrt{\left(\frac{GM}{c^{2}}\right)^{2}\left(1+\lambda^{2}\right)^{2}-a^{2}}\label{2.126}
\end{eqnarray}
respectively.

\begin{figure}[!tbp]
  \begin{subfigure}[b]{0.25\textwidth}
    \includegraphics[width=\textwidth]{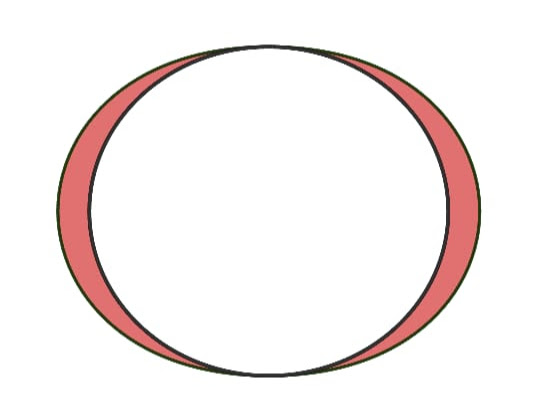}
    \caption{$\lambda=0$.}
    \label{fig:f1}
  \end{subfigure}
  \hfill
  \begin{subfigure}[b]{0.25\textwidth}
    \includegraphics[width=\textwidth]{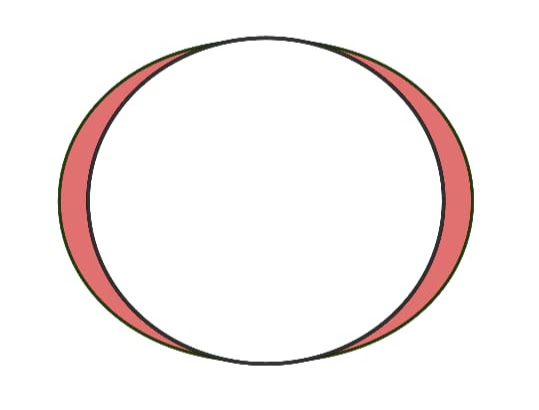}
    \caption{$\lambda=1/4\lambda_{\mathrm{crit}}$.}
    \label{fig:f2}
  \end{subfigure}
  \hfill
  \begin{subfigure}[b]{0.25\textwidth}
    \includegraphics[width=\textwidth]{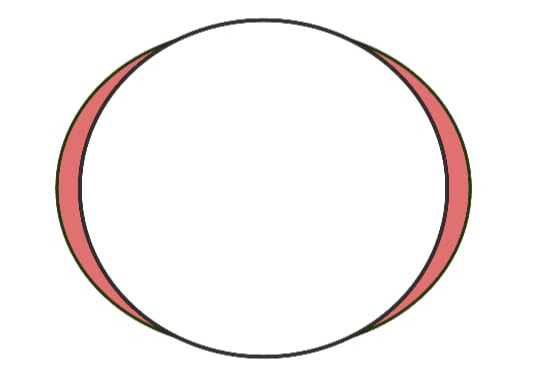}
    \caption{$\lambda=1/2\lambda_{\mathrm{crit}}$.}
    \label{fig:f3}
  \end{subfigure}
  \hfill
  \begin{subfigure}[b]{0.25\textwidth}
    \includegraphics[width=\textwidth]{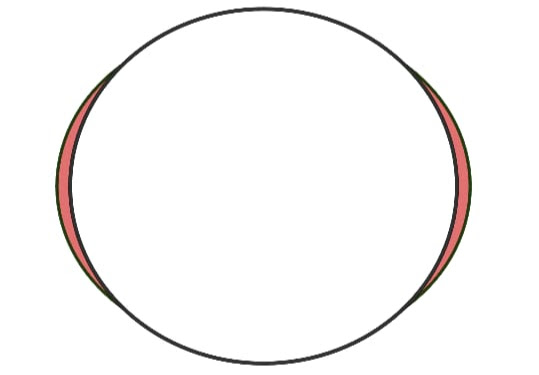}
    \caption{$\lambda=3/4\lambda_{\mathrm{crit}}$.}
    \label{fig:f4}
  \end{subfigure}
  \hfill
  \begin{subfigure}[b]{0.25\textwidth}
    \includegraphics[width=\textwidth]{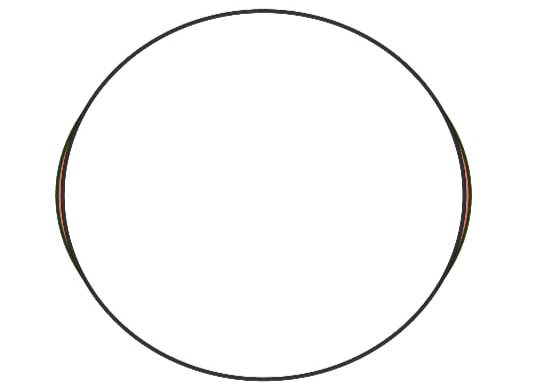}
    \caption{$\lambda=9/10\lambda_{\mathrm{crit}}$.}
    \label{fig:f5}
  \end{subfigure}
  \hfill
  \begin{subfigure}[b]{0.25\textwidth}
    \includegraphics[width=\textwidth]{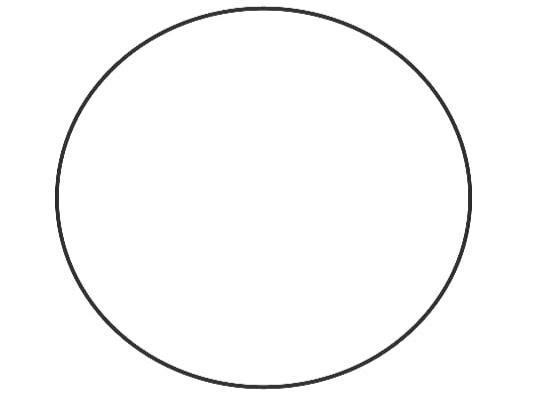}
    \caption{$\lambda=\lambda_{\mathrm{crit}}$.}
    \label{fig:f6}
  \end{subfigure}
  \caption{Schematic diagram of the throat and ergoregion of the DSRW for different values of the deformation parameter. The throat is the white region with black rim and the ergoregion is the red section with black rim.}
  \label{fig:lambda}
\end{figure}

In Figure \ref{fig:lambda} we show a representation of the wormhole ergosurface and throat for fixed values of $a$ and $M$, varying the parameter $\lambda$. For the case $\lambda=0$, the line element of the wormhole coincides with Kerr's: we see that the entire throat is contained by the ergosphere; both surfaces coincide at the poles. If we increase the value of the parameter $\lambda$, while the wormhole ergosphere is unchanged, the throat increases in size, so the ergosphere no longer contains it completely.

The graphs in Figure \ref{fig:lambda} reveal two features: 

\begin{itemize}

\item There exists a value of $\lambda$, which we denote $\lambda_{\mathrm{crit}}$ such that the ergosphere is completely contained in the throat of the wormhole, and both surfaces coincide in the equator of the wormhole. Then, for all values of the deformation parameter $\lambda>\lambda_{\mathrm{crit}}$, the wormhole has no ergoregion (if $\lambda=\lambda_{\mathrm{crit}}$, the ergoregion and the throat coincide only in the equatorial plane of the wormhole). Mathematically, this means
\begin{eqnarray} 
r_{+}(\lambda_{\mathrm{crit}})&=&r_{S^{+}}\\
\frac{GM}{c^{2}}(1+\lambda_{\mathrm{crit}}^{2})+\sqrt{(\frac{GM}{c^{2}})^{2}(1+\lambda_{\mathrm{crit}}^{2})^{2}-a^{2}}&=& \frac{GM}{c^{2}}+\sqrt{(\frac{GM}{c^{2}})^{2}-a^{2}\cos^{2}{\frac{\pi}{2}}}.
\label{2.112}
\end{eqnarray}

From this we obtain that:

\begin{equation}
    \lambda_{\mathrm{crit}}=\frac{ac^{2}}{2GM}.
\end{equation}

\item It is also of interest to find  the angle of intersection of the two surfaces, $\theta_{\mathrm{crit}}$, and which exists as long as the deformation parameter does not exceed the deformation parameter $\lambda_{\mathrm{crit}}=ac^{2}/(2GM)$. To find it, we equal expressions \eqref{2.124} and \eqref{2.126}

\begin{align}
    r_{+}&=r_{S^{+}}, \\
    \frac{GM}{c^{2}}(1+\lambda^{2})+\sqrt{\left(\frac{GM}{c^{2}}\right)^{2}(1+\lambda^{2})^{2}-a^{2}}&=\frac{GM}{c^{2}}+\sqrt{\left(\frac{GM}{c^{2}}\right)^{2}-a^{2}\cos^{2}{\theta_{\mathrm{crit}}}}.
\end{align}

Finally, we obtain

\begin{equation}
    \cos^{2}{\theta_{\mathrm{crit}}}=1-\frac{2GM}{c^{2}a^{2}}\left(\lambda^{2}\sqrt{\left(\frac{GM}{c^{2}}\right)^{2}(1+\lambda^{2})^{2}-a^{2}}+\frac{GM}{c^{2}}\lambda^{2}(1+\lambda^{2})\right).
\end{equation}

\end{itemize}

At $\lambda=0$, we obtain $\theta_{\mathrm{crit}}=0$, recovering the expected result for a Kerr black hole (the surfaces coincide at the poles). On the other hand, if we evaluate the angle of intersection in $\lambda=ac^{2}/2GM$, we get $\theta=\pi/2$: the surfaces coincide in the equatorial plane, and the ergosphere completely surrounds the throat, so that the wormhole does not present an ergoregion.

It can also be shown that the DSRW spacetime has stable circular orbits for massive particles, and in particular, a stable circular orbit of minimum radius called ISCO. Its expression is independent of the deformation parameter $\lambda$, and thus coincides with that for a Kerr black hole \cite{2019Karimov}.

\subsection{Energy conditions}

Transversable wormholes violate the Null Energy Conditions (NEC) $T_{\mu \nu} k^{\mu} k^{\nu} \ge 0$, where $k^{\mu}$ is a null vector \citep{mor+88,vis95}. In what follows, we will analyze the conditions under which the NEC is satisfied.

We consider an imperfect fluid with an energy-momentum tensor given by the expression

\begin{equation}
    T^{ab}=\rho\hat{\epsilon}^{a}_{0}\hat{\epsilon}^{b}_{0}+p_{\mathrm{r}}\hat{\epsilon}^{a}_{1}\hat{\epsilon}^{b}_{1}+p_{\mathrm{\theta}}\hat{\epsilon}^{a}_{2}\hat{\epsilon}^{b}_{2}+p_{\mathrm{\phi}}\hat{\epsilon}^{a}_{3}\hat{\epsilon}^{b}_{3}+q\hat{\epsilon}^{a}_{0}\hat{\epsilon^{b}}_{3}+q\hat{\epsilon}^{a}_{3}\hat{\epsilon}^{b}_{0},
\end{equation}

where $\rho$ denotes the density, $p_{\mathrm{r}}$, $p_{\mathrm{\theta}}$, $p_{\mathrm{\phi}}$ the principal pressures and $q$ is the heat flux. The vectors $\hat{\epsilon^{a}_{\alpha}}$ form an orthonormal basis that satisfies the conditions 
\begin{equation}
g_{ab}\hat{\epsilon}^{a}_{\alpha}\hat{\epsilon}^{b}_{\beta}=\eta_{\alpha\beta},
\end{equation}
with $\eta_{\alpha}=\mathrm{diag}(-1,1,1,1,1)$, the Minkowski metric. The null vector $k^{\alpha}$ can be written as
\begin{equation}
k^{\alpha}=\hat{\epsilon^{\alpha}_{0}}+a\hat{\epsilon^{\alpha}_{1}}+b\hat{\epsilon^{\alpha}_{2}}+c\hat{\epsilon^{\alpha}_{3}},
\end{equation}
where $a$, $b$ and $c$ are arbitrary functions of the coordinates that fulfill the condition $a^{2}+b^{2}+c^{2}=1$. After multiple algebraic operations, the NEC implies

\begin{equation}
    \rho-2cq+a^{2}p_{\mathrm{r}}+b^{2}p_{\mathrm{\theta}}+c^{2}p_{\mathrm{\phi}}\left.|\right._{r_{+}}\geq0.
\end{equation}

If we take $b=c=0$, then $a=1$ and we get

\begin{equation}
    \rho+p_{\mathrm{r}}\left.|\right._{r_{+}}\geq0.
    \label{2.5}
\end{equation}

If $a=c=0$, then $b=1$ and this yields

\begin{equation}
    \rho+p_{\mathrm{\theta}}\left.|\right._{r_{+}}\geq0.
    \label{2.6}
\end{equation}

Finally, if $a=b=0$, then $c=1$ and 

\begin{equation}
    \rho+p_{\mathrm{\phi}}-2q\left.|\right._{r_{+}}\geq0.
    \label{2.7}
\end{equation}

Then, the NEC expressed in terms of the density $\rho$, principal pressures $p_{\mathrm{r}}$, $p_{\mathrm{\theta}}$, $p_{\mathrm{\phi}}$ and heat flux $q$ is satisfied if

\begin{equation}
    \rho+p_{\mathrm{r}}\left.\right|_{r_{+}}\geq0, ~~ \rho+p_{\mathrm{\theta}}\left.\right|_{r_{+}}\geq0, ~~ \rho+p_{\mathrm{\phi}}-2q\left.\right|_{r_{+}}\geq0.
\end{equation}

From the line element \eqref{4.1}, we can define the Einstein tensor in the orthonormal coordinate basis, whose non-zero components are:

\begin{eqnarray}
    G_{\hat{t}\hat{t}}&=&-\frac{g_{\phi\phi}}{\mathrm{det}}G_{tt}, \\
    G_{\hat{t}\hat{\phi}}&=&G_{\hat{\phi}\hat{t}}=-\left(G_{t\phi}+G_{\phi\phi}\frac{g_{t\phi}}{g_{\phi\phi}}\right)\sqrt{-\frac{1}{\mathrm{det}}},\\
    G_{\hat{r}\hat{r}}&=&\frac{1}{g_{rr}}G_{rr},\\
    G_{\hat{\theta}\hat{\theta}}&=&\frac{1}{g_{\theta\theta}}G_{\theta\theta},\\
    G_{\hat{\phi}\hat{\phi}}&=&\frac{1}{g_{\phi\phi}}G_{\phi\phi},
\end{eqnarray}

with $\mathrm{det}$ being the determinant of the metric, $\mathrm{det}=g_{tt}g_{\phi\phi}-g_{t\phi}^{2}$. Then we get the expressions for $\rho$, $p_{\mathrm{r}}$, $p_{\mathrm{\theta}}$, $p_{\mathrm{\phi}}$ and $q$ from Einstein's field equations,

\begin{eqnarray}
    \rho(r,\theta)&=&-\frac{c^{4}}{8\pi G}\frac{\mathrm{det}}{g_{tt}g_{\phi\phi}}G_{\hat{t}\hat{t}}, \label{2.14}\\
    p_{\mathrm{r}}(r,\theta)&=&\frac{c^{4}}{8\pi G}G_{\hat{r}\hat{r}}, \label{2.15}\\
    p_{\mathrm{\theta}}(r,\theta)&=&\frac{c^{4}}{8\pi G}G_{\hat{\theta}\hat{\theta}}, \label{2.16}\\
    p_{\mathrm{\phi}}(r,\theta)&=&\frac{c^{4}}{8\pi G}G_{\hat{\phi}\hat{\phi}}, \label{2.17}\\
    q(r,\theta)&=&-\frac{c^{4}}{8\pi G}\left(G_{\hat{t}\hat{\phi}}-G_{\hat{t}\hat{t}}\frac{g_{t\phi}^{2}}{g_{tt}g_{\phi\phi}}\right)\frac{\sqrt{-\mathrm{det}}}{g_{t\phi}}. \label{2.18}
\end{eqnarray}

It's important to note that, if we set the deformation parameter $\lambda=0$ in expressions \eqref{2.14}-\eqref{2.18}, they are all zero. This is the expected result, since in the limit $\lambda = 0$ we recover the Kerr metric, which is a vacuum solution of Einstein's field equations.

For a further analysis of NEC, we considered a $10\,M_{\odot}$ wormhole and took three values for the wormhole spin: $a=0.1,\,0.5,\,0.95$, expressed in units of mass ($a/(GM/c^{2})$).

\begin{figure}
    \centering
    \includegraphics[width=0.9\textwidth] {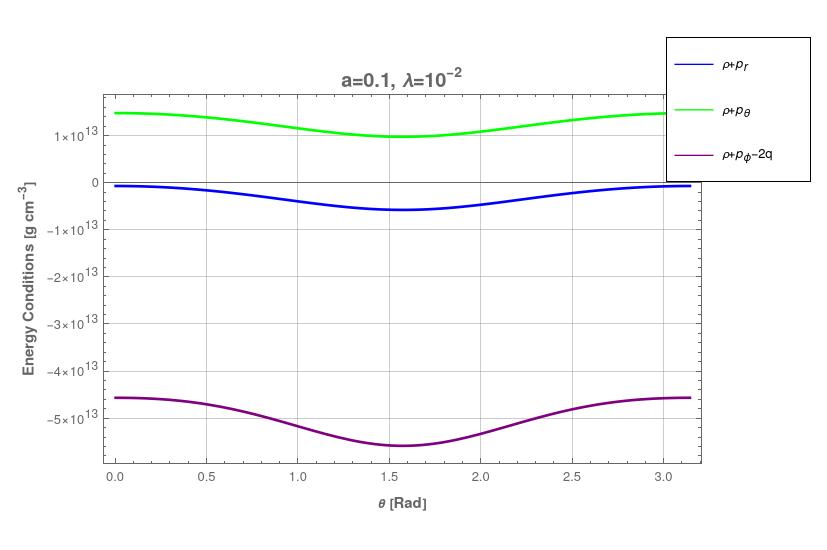}
    \caption{NEC for a $2\,M_{\odot}$ wormhole with a spin parameter $a=0.1$ and a deformation parameter $\lambda=10^{-2}$.}
  \label{fig:NEC1}
\end{figure}

\begin{figure}
    \centering
    \includegraphics[width=0.9\textwidth] {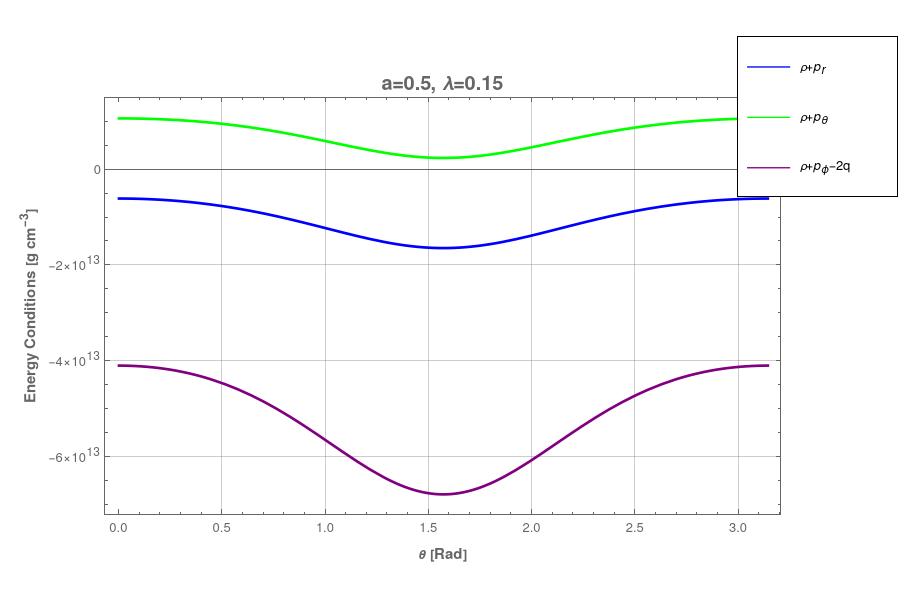}
    \caption{NEC for a $2\,M_{\odot}$ wormhole with a spin parameter $a=0.5$ and a deformation parameter $\lambda=0.15$.}
  \label{fig:NEC2}
\end{figure}

\begin{figure}
    \centering
    \includegraphics[width=0.9\textwidth] {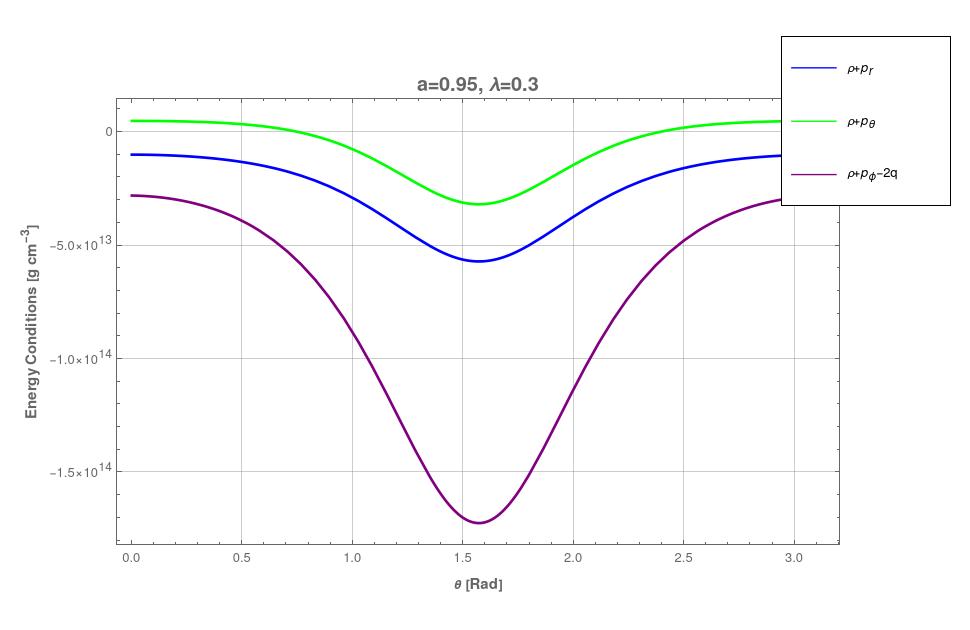}
    \caption{NEC for a $2\,M_{\odot}$ wormhole with a spin parameter $a=0.95$ and a deformation parameter $\lambda=0.3$.}
  \label{fig:NEC3}
\end{figure}

The Figures \ref{fig:NEC1}, \ref{fig:NEC2} and \ref{fig:NEC3} show the NEC at the throat with the chosen spin parameters $a=0.1, 0.5, 0.95$ and the deformation parameters $\lambda=10^{-2}, 0.15, 0.3$ respectively.

It can be seen from the three figures that conditions \eqref{2.5} and \eqref{2.7} are globally violated, and that condition \eqref{2.6} is violated only locally close to the equator (see  Figure \ref{fig:NEC3}; this is due to the choice of the deformation parameter, as we will explain next.

In the expression \eqref{2.14} there is a divergence due to the term $1/g_{tt}$, that takes the form $r_{+}(-2GM+c^{2}r_{+})+a^{2}c^{2}\cos^{2}{\theta}=0$. From it, we can define a value of the deformation parameter (we will denote it as $\lambda_{\mathrm{div}}$) such that if we choose values of the deformation parameter in the range $0<\lambda\leq\lambda_{\mathrm{div}}$, \eqref{2.14} presents two divergences in the axial angle, which are symmetric with respect to the equator of the wormhole. In turn, due to the presence of this divergence, there is a local violation of the NEC conditions \eqref{2.5}-\eqref{2.7} in regions close to the equator. 

For this reason, we consider that the deformation parameter takes the range of values of $\lambda_{\mathrm{div}}<\lambda\leq1$, where no divergence is appreciated. Under these conditions, NEC \eqref{2.5} and \eqref{2.7} are violated globally, regardless of the assumed value for the spin parameter $a$, as can be seen in Figures \ref{fig:NEC1}, \ref{fig:NEC2} and \ref{fig:NEC3}. In contrast, we find that \eqref{2.6} is at most locally violated in a neighborhood of the wormhole equator; we can define a value of the deformation parameter (we will denote it as $\lambda_{\mathrm{NEC}2}$) such that for $\lambda_{\mathrm{div}}<\lambda\leq\lambda_{\mathrm{NEC}2}$, \eqref{2.6} is locally violated near the equator, and for  $\lambda_{\mathrm{NEC}2}<\lambda\leq1$, \eqref{2.6} is globally satisfied.

The quantities $\lambda_{\mathrm{div}}$ and $\lambda_{\mathrm{NEC}2}$ have been calculated for the chosen spin parameters and are shown in Table \ref{table:lambdaparameters}. From the table it can be seen that both deformation parameters increase with the spin parameter and that for the case of spin a=0.1 no local violation of \eqref{2.6} is observed considering the range of deformation parameters $\lambda_{\mathrm{div}}<\lambda\leq1$. Also, by taking the spin parameter $a=0.95$ and the deformation parameter $\lambda=0.3$, we see that the latter is in the interval $\left(\lambda_{\mathrm{div}},\lambda_{\mathrm{NEC2}}\right)$, and due to this choice, as we said before, the condition \eqref{2.6} is locally violated.

\begin{table}
\large
\centering
\begin{tabular}{c | c | c} 
  & $\lambda_{\mathrm{div}}$ & $\lambda_{\mathrm{NEC}2}$\\ [0.5ex] 
 \hline\hline
 $a=0.1 \;GM/c^2$ & 0.00250 & - \\ 
 \hline
 $a=0.5 \;GM/c^2$ & 0.06275 & 0.1329 \\
 \hline
 $a=0.95 \;GM/c^2$ & 0.2393 & 0.4137 \\
 \hline
\end{tabular}
\caption{Deformation parameters $\lambda_{\mathrm{div}}$ and $\lambda_{\mathrm{NEC}2}$ for DSRW with mass $M=10\, M_{\odot}$ and spin parameter $a=0.1,\,0.5,\,0.95$.}
\label{table:lambdaparameters}
\end{table}

\section{Blandford-Znajek mechanism in rotating wormholes}

The Blanford-Znajek mechanism \citep{1977Blandford} is now widely used to refer to any process of extracting rotational energy from a black hole by means of an electromagnetic field \citep{rom+14}.


By its very nature, the event horizon of a black hole is causally "disconnected" from the electromagnetic outflow. Hence, the presence of the event horizon is not a condition necessary for this process to occur. The ergosphere, together with the presence of a magnetic field, are the main components that would allow the extraction of rotational energy from the compact object. 

The latter implies that if a compact object has an ergosphere and is surrounded by a magnetic field, the Blandford-Znajek mechanism could also take place. This is exactly the case for the rotating wormhole discussed in the previous section.

Consider a magnetosphere resulting from the flow of currents in an accretion disk across the equator of the compact object. Assume that it is in the force-free regime (the inertia of matter outside the accretion disk is neglected, but the charge density is high enough to shield the component of the electric field that is parallel to the magnetic field); so, the magnetosphere is such that the Lorentz force is zero and the magnetic pressure dominates over the astrophysical plasma pressure. The dynamics of the magnetosphere is then described by Maxwell's equations

\begin{equation}
    \nabla_{\mu}F^{\mu\nu}=j^{\nu},  ~~\nabla_{[\rho}F_{\mu\nu]}=0,
    \label{3.6}
\end{equation}

under the force-free

\begin{equation}
    F_{\mu \nu}J^{\nu}=0,
\end{equation}

and MHD ideal condition

\begin{equation}
    F^{* \mu \nu}F_{\mu \nu}=0.
    \label{eq:contraccion}
\end{equation}

Following these considerations, the electromagnetic tensor takes the form 

\begin{equation}
F_{\mu \nu}= \sqrt{-g} \begin{pmatrix}
0 & -\omega B^{\theta} & \omega B^{r} & 0 \\
\omega B^{\theta} & 0 & B^{\phi} & -B^{\theta} \\
-\omega B^{r} & -B^{\phi} & 0 & B^{r} \\
0 & B^{\theta} & -B^{r} & 0
\end{pmatrix},
\label{3.32}
\end{equation}

where $\omega(r,\theta)=-A_{t,\theta}/A_{\phi,r}=-A_{t,r}/A_{\phi,\theta}$ is the magnetic frecuency, $g=-c^{2}\Sigma(r,\theta)^{2}\sin^{2}{\theta}\Delta/\hat{\Delta}$ is the determinant of the metric and $B^{r}$, $B^{\theta}$ and $B^{\phi}$ are the magnetic field components.

In this regime, the matter component of the energy-momentum tensor is overshadowed by the electromagnetic component, so we will only consider the latter, 

\begin{equation}
    T^{\mu}_{\nu}=F_{\nu \lambda}F^{\mu \lambda}+\frac{1}{4}\delta^{\mu}_{\nu}F^{\kappa \lambda}F_{\kappa \lambda}.
\end{equation}

From the conservation of the energy-momentum tensor $\nabla_{\mu} T^{\mu}_{\nu} = 0$, follows that $\mathcal{E}^{\nu} = T^{\nu}_{t}$ is conserved:  an electromagnetic energy flow in radial direction can be defined \cite{1977Blandford, 2021Konoplya, 1973Misner}

\begin{equation}
    \mathcal{E}^{r}=-T^{r}_{t}=\left(F_{t \theta}F_{\theta \phi}g^{r \phi}-F_{r \theta}F_{t \theta}g^{rr}-F_{t \theta}^{2}g^{t r}\right) g^{\theta \theta}.
    \label{3.46}
\end{equation}

In order to compute this quantity, it is necessary to know the geometry of the magnetic field. This is the topic of the next section.

\subsection{A model for the geometry of the magnetic field}



To compute the geometry of the magnetic field in the background spacetime, it is usual to take the following assumptions: due to the stationary and axisymmetric nature of spacetime, it is imposed that the magnetosphere behaves in the same way; so, a gauge is chosen such that the quadrivector potential is independent of the temporal and azimuthal coordinates
$\partial_{t}A_{\mu} = \partial_{\phi}A_{\mu} =0. $
Then, three functions denoted $\Psi(r,\theta)$, $\Omega$ and $I$ are defined: 
\begin{equation}
    \Psi(r,\theta)\equiv A_{\phi}(r,\theta),
\end{equation}
which represents the magnetic flux through a circular loop of radius $r\,\sin{\theta}$ surrounding the polar axis of the BH. The angular velocity of the magnetic field lines takes the form
\begin{equation}
    \Omega(r,\theta)=-\frac{\partial_{r}A_{t}}{\partial_{r}\Psi}=-\frac{\partial_{\theta}A_{t}}{\partial_{\theta}\Psi},
\end{equation}
Finally,  $I$ represents the poloidal current
\begin{equation}
    I=\sqrt{-g}F^{\theta r},
\end{equation}
with $g$ the determinant of the line element. Notice that $\Omega$ and $I$ are functions that depend only on the magnetic flux, that is, $\Omega=\Omega(\Psi)$ and  $I = I(\Psi)$ \citep{2022Camilloni}.

Once this is established, the geometry of the magnetic field is obtained from the solution of the partial derivative equation called stream-equation,

\begin{equation}
    \Omega\partial_{\rho}\left(\sqrt{-g}F^{t\rho }\right)-\partial_{\rho}\left(\sqrt{-g}F^{\phi\rho}\right)+F_{r\theta}\frac{dI}{d\Psi}=0,
    \label{3.12}
\end{equation}
whose unknown variable is the magnetic flux $\Psi(r,\theta)$.

Solving the latter equation is extremely complicated. In their seminal paper, Blandford and Znajek followed a perturbative approach to obtain a solution in Kerr spacetime \citep{1977Blandford,2010ApJ...711...50T,2010ApJ...711...50T,2014PhRvD..90l4009Z,2008PhRvD..78b4004T}

In this paper, however, we take a different approach. Since our mail goal is to determine whether rotational energy can be extracted in the form of electromagnetic flux in an axisymmetric wormhole spacetime, as a first approach to the problem we assume a particular geometry for the magnetic field.

 We consider that in the region outside the ergosphere, the magnetic field is asymptotically uniform and points in the direction of axial symmetry of the black hole \cite{1974Wald}, i.e.

\begin{equation}
    \Vec{B}=B_{0}\hat{z}, \quad r\rightarrow\infty.
\end{equation}

\begin{figure}
    \centering
    \includegraphics[width=0.7\textwidth]{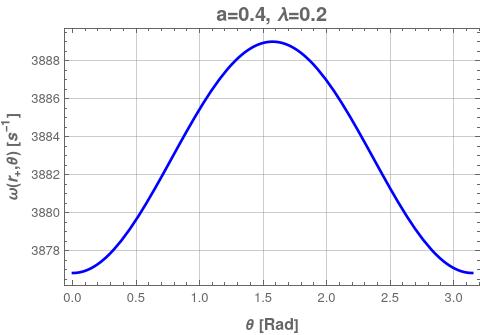}
    \caption{Magnetic frequency over the throat for a $2\,M_{\odot}$ wormhole with a spin parameter $a=0.4$ and a deformation parameter $\lambda=0.2$.}
  \label{fig:omega}
\end{figure}

From this it is found that the magnetic field has non-zero components in the radial and axial direction (for further details, see Section \ref{appendix})

\begin{equation}
    \Vec{B}=B^{r}(r,\theta)\hat{r}+r_{\mathrm{g}}B^{\theta}(r,\theta)\hat{\theta}
\end{equation}

with 

\begin{eqnarray}
    B^{r}(r,\theta)&=&\frac{1}{\sqrt{-g}}F_{\theta\phi}=\frac{B_{0}}{2\Sigma\sin{\theta}}\sqrt{\frac{\hat{\Delta}}{\Delta}}\frac{\partial X}{\partial\theta}, \\
    B^{\theta}(r,\theta)&=&\frac{1}{\sqrt{-g}}F_{\phi r}=-\frac{B_{0}}{2\Sigma\sin{\theta}}\sqrt{\frac{\hat{\Delta}}{\Delta}}\frac{\partial X}{\partial r}, \\
    X&=&g_{\phi\phi}+2\frac{a}{c}g_{t\phi}=\frac{\sin^{2}{\theta}}{\Sigma}\left(\rho^{2}-4\frac{GM}{c^{2}}a^{2}r\right), \\
    \rho^{2}&=&(r^{2}+a^{2})^{2}-a^{2}\Delta\sin^{2}{\theta},\\
    \Delta&=&r^{2}-2GM/c^{2}r+a^{2},
\end{eqnarray}

and, from the definition of the electromagnetic tensor, the frequency of the magnetic field is 

\begin{eqnarray}
    \omega&=&\frac{F_{t\theta}}{F_{\theta\phi}}=-\frac{U_{,\theta}}{X_{,\theta}},\\
    U&=&g_{t\phi}+2\frac{a}{c}g_{tt}=-2ac\left[1-\frac{2GMr\cos^{2}{\theta}}{c^{2}\Sigma}\right].
\end{eqnarray}

whose dependence on the polar coordinate is contained in the function $\Sigma$. This expression shows that the magnetic frequency is well-behaved at the throat and has a maximum at the equator of the wormhole, i.e. $\theta=\pi/2$, and two minima at the poles, $\theta=0,\pi$, as it can be seen in Figure \ref{fig:omega}. Furthermore, in the limiting case $\lambda=0$ (Kerr black hole) it is independent of the $\theta$ coordinate, since $\Delta\big|_{r_{EH}}=0$,

\begin{equation}
    \omega\big|_{r_{EH}}=\frac{ac^3}{2GM\sqrt{-a^2+\left(\frac{GM}{c^2}\right)^{2}}}
\end{equation}

We assume that, due to the frame-dragging phenomenon, the magnetic field lines penetrating the ergosphere develop an azimuthal component $B^{\phi}$ so that it cancels over the ergosphere and is maximal over the throat. At the same time, the corresponding radial and axial components are modified so that they cancel over the throat and coincide with the external magnetic field when evaluated at the ergosphere. With these considerations, the magnetic field is described as

\begin{equation}
    \begin{split}
        \Vec{B}_{\mathrm{in}}(r,\theta) & =\Vec{B}_{\mathrm{new}}+r_{\mathrm{g}}B^{\phi}(r,\theta)\hat{\phi}, \\
        \Vec{B}_{\mathrm{new}}(r,\theta) & =B^{r}_{\mathrm{new}}\hat{r}+r_{\mathrm{g}}B^{\theta}_{\mathrm{new}}=\left(B^{r}(r,\theta)\hat{r}+r_{\mathrm{g}}B^{\theta}(r,\theta)\hat{\theta}\right)y(r,\theta)
    \end{split}
\end{equation}

where the subscript \textit{in} indicates that it is the magnetic field within the ergoregion and the subscript \textit{new} indicates that it is a new poloidal component. By $y(r,\theta)$ we mean a linear function such that when evaluated over the ergosphere  $y (r_{S^{+}}, \theta) =1$ and when evaluated over the throat it cancels out,

\begin{equation}
    y(r,\theta)=\frac{r-r_{+}}{r_{S^{+}}-r_{+}}.
\end{equation}

Given these considerations, we need to obtain an explicit form for the $B^{\phi}$ component, and for this purpose we will impose one last condition: The magnitude of the magnetic field inside the ergosphere will be uniform and coincide with the magnitude of the magnetic field above the ergosphere. Then, the azimuthal component of the magnetic field $B^{\phi}(r,\theta)$ is

\begin{equation}
    B^{\phi}(r,\theta)=-\frac{1}{r_{\mathrm{g}}}\sqrt{B(r_{S^{+}},\theta)^{2}-y(r,\theta)^{2}\left(B^{r}(r,\theta)^{2}+r_{\mathrm{g}}^{2}B^{\theta}(r,\theta)^{2}\right)},
\end{equation}

where the choice of the sign is made in order to obtain positive rates of energy extraction.

In this work, the accretion disk is present only to provide a source that allows the existence of a large magnetic field in the wormhole environment. A possible mechanism to generate the magnetic field in the disk is the dynamo effect: the resulting geometry of the magnetic field in the disk has a dominant poloidal component, as shown in the recent 3D GRMHD simulations by Jacquemin-Ide and collaborators \citep{jac+24}. As explained above, the initial magnetic field in our work (outside the ergosphere) has only poloidal components. Once the magnetic field enters the ergoregion, it develops a toroidal component due to frame-drag phenomena, as required by the Blandford-Znajek mechanism.

\section{Results}

\subsection{Parameters for the spacetime model}
In view of the proposed model for the geometry of the magnetic field, we are now going to choose the range of values for the spin and deformation parameters such that a Poynting flux is possible. In what follows, we take geometrized units ($G=c=M=1$), so that the spin parameter takes values in the range $0 \leq a\leq 1$.


We showed in Section \eqref{sec:2} that if the deformation parameter exceeded the value $\lambda_{\mathrm{crit}}$, the spacetime did not have an ergosphere, which we identify as necessary  for the emission of an electromagnetic flux. Thus, as a first condition, we obtain that the deformation parameter takes values in the range $0 \leq \lambda \leq \lambda_{\mathrm{crit}}=a/2$. 

On the other hand, we assume that the magnetosphere is produced by charges in an accretion disk around the wormhole. The accretion disk lies in the equatorial plane and its innermost radius corresponds to the radial coordinate of the last stable circular orbit (or ISCO ``innermost stable circular orbit'')  ($r_{\mathrm{disk}}\geq r_{\mathrm{ISCO}}$). This assumption regarding the inner boundary of the disk is justified in the optically thick and geometrically thin disk models of \cite{Shakura1973A&A&A....24..337S} and its relativistic version of \cite{Novikov1973blho.conf..343N} and \cite{Page1974ApJ...191..499P}, that are used when the accretion onto the compact object is subcritical, that is $\dot{M} < \dot{M}_{\mathrm{crit}}$, with $\dot{M}_{\mathrm{crit}} = L_{\mathrm{Edd}}/c^2$, where $L_{\mathrm{Edd}}$ denotes the Eddington luminosity\footnote{Depending on the mass accretion rate, different accretion regimes are possible. At very low rates $\dot{M} \ll \dot{M}_{\mathrm{crit}}$, the disk truncates at a certain distance and the inner region is filled with a two-temperature plasma. This geometrically thick and optically thin hot accretion flow is more efficient than thin disks in launching and collimating relativistic jets. As discussed by \citep{rom+20,rom21}, the jet is initiated as a Poynting flux emerging from the black hole ergosphere. The neutral particles around the hole could be responsible for the mass loading of the jet through various processes. At the opposite extreme, the accretion is super-Eddington and photon trapping in the disk becomes important. The disk becomes optically and geometrically thick. Strong winds are driven out of the disk by the intense radiation pressure. This wind helps to collimate the Poynting flux and the inner jet, allowing the jet to propagate over large distances. In each case, the properties of the accreted flow could contribute to the launching, collimation, and mass loading of astrophysical jets. However, since in our work we are only concerned with the generation of an electromagnetic flux via the Blandford-Znajek mechanism, and the latter depends on the background spacetime and the geometry of the magnetic field, we expect that a different accretion model will not significantly change our predictions.}. Part of the accretion disk must be contained within the ergosphere. This condition is expressed mathematically as
\begin{equation}
    r_{\mathrm{ISCO}}\leq r_{S^{+}}(\theta=\pi/2)=2M=2.
    \label{4.4}
\end{equation}

\begin{figure}
    \centering
    \includegraphics[width=0.6\textwidth]{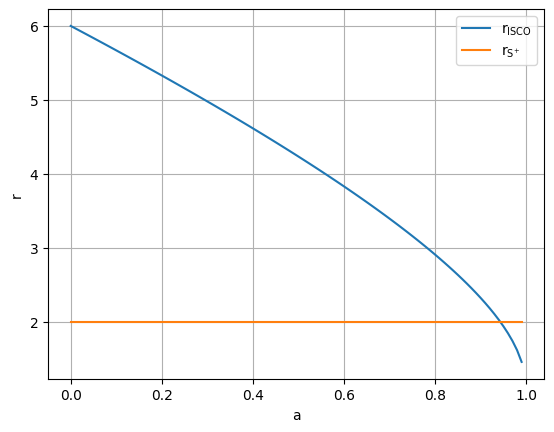}
    \caption{Radius of the ISCO and ergoregion over the equator as a function of spin parameter $a$.}
  \label{fig:risco}
\end{figure}

In Figure \ref{fig:risco}, we plot the radial coordinate of the ISCO and the radius of the ergosphere at the equator as a function of the spin $a$. From the Figure we deduce that the condition \eqref{4.4} is satisfied only if the spin parameter is in the range $0.94281\leq a\leq 1$, i.e., only very fast rotating Damour-Solodukhin wormholes.

But this condition is not sufficient, since we also require that the minimum radius of the accretion disk be larger than the throat of the wormhole. This is achieved by taking those values of the deformation parameter $\lambda$ such that the ISCO is outside the throat.

\begin{figure}
    \centering
    \includegraphics[width=0.6\textwidth] {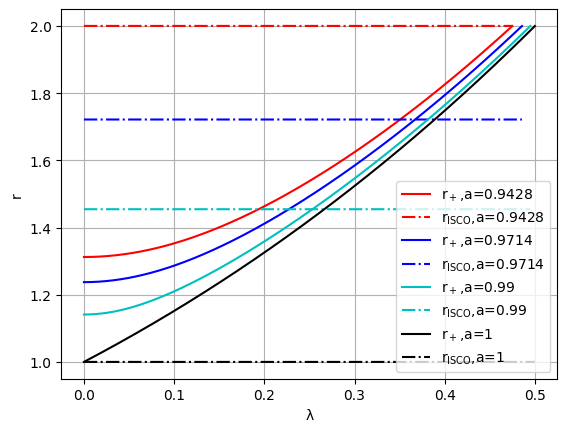}
    \caption{Radius of the last stable circular orbit and throat over the equator for three different values of $a$.}
  \label{fig:lambdarisk}
\end{figure}

In Figure \ref{fig:lambdarisk} we plot with solid line the throat radius (solid line) as a function of the deformation parameter $\lambda$ for four different values of spin: $a = 0.9428$ (red), $a = 0.95$ (blue), $a = 0.9714$ (cyan) and $a = 1$ (black). The four dashed curves correspond to the radius of ISCO for each of these spin values. Since $r_{\mathrm{ISCO}}$ does not depend on $\lambda$ (they take the same values as for a Kerr black hole), the curves are straight lines parallel to the abscissa axis. 

We  observe that for the minimum value of $a$, no matter what value we assign to the deformation parameter (remember that $0\leq\lambda\leq\lambda_{\mathrm{crit}}=a/2$, where we have already normalized $M=1$), the ISCO is outside the throat. As we increase the value of the spin, both the radius of the throat and the radius of the ISCO decrease; we see that there is a value of the deformation parameter, denoted $\tilde{\lambda}_{\mathrm{crit}}$, where the innermost stable circular orbit coincides with the throat, $r_{\mathrm{ISCO}}=r_{+}$. For $\lambda \geq \tilde{\lambda}_{\mathrm{crit}}$, the last stable circular orbit falls inside the throat, violating the assumptions of our model. 

When the limiting value of the spin parameter, $a=1$, is reached, if $\lambda>0$, then $r_{+}>r_{\mathrm{ISCO}}$, and therefore, the ISCO would fall inside the throat, matching it for $\lambda = 0$. But there is a problem with this situation: for $\lambda=0$, the metric of the wormhole reduces to that of a Kerr black hole. Moreover, when the spin is $a=1$, the event horizons vanish leaving room for a naked singularity \citep{2006Hobson}.

From this analysis we conclude that to model a rotating wormhole presenting an accretion disk inside the ergosphere, we have to take values of the spin parameter in the interval $0.94281GM/c^{2}\leq a<GM/c^{2}$, having recovered the constants so that the spin parameter has units of length. 

Once the spin parameter $a$ has been fixed, the deformation parameter $\lambda$ 
can only take values in the range $0\leq\lambda\leq\Tilde{\lambda}_{\mathrm{crit}}\leq\lambda_{\mathrm{crit}}$, with $\Tilde{\lambda}_{\mathrm{crit}}$ dependent on the choice of the spin parameter $a$.

\subsection{Poynting Flux}

Now, we have all the conditions and quantities necessary to analyze the electromagnetic flux emission from a wormhole. 

Using the definition of the electromagnetic tensor \eqref{3.32} and taking into account the form of  the determinant of the metric tensor  $-g=c^{2}\Sigma^{2}\sin^{2}{\theta}\Delta/\hat{\Delta}$, the expression for  $\mathcal{E}^{r}$ \eqref{3.46} reduces to

\begin{equation}
    \mathcal{E}^{r}=-c^{2}\omega(r,\theta)B^{r}_{\mathrm{new}}(r,\theta)B^{\phi}(r,\theta)\Delta\sin^{2}{\theta}.
    \label{52}
\end{equation}

\begin{figure}
    \centering
    \includegraphics[width=0.9\textwidth] {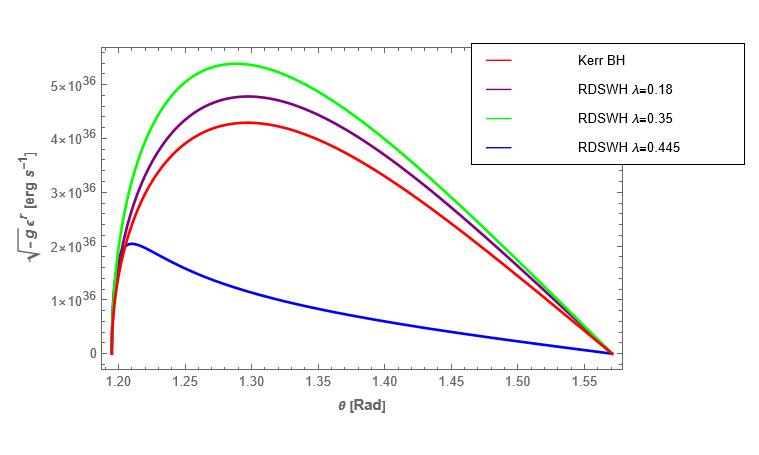}
    \caption{Electromagnetic energy flux per unit solid angle for spin parameter $a=0.95$, varying the deformation parameter.}
  \label{fig:a095}
\end{figure}

\begin{figure}
    \centering
    \includegraphics[width=0.9\textwidth] {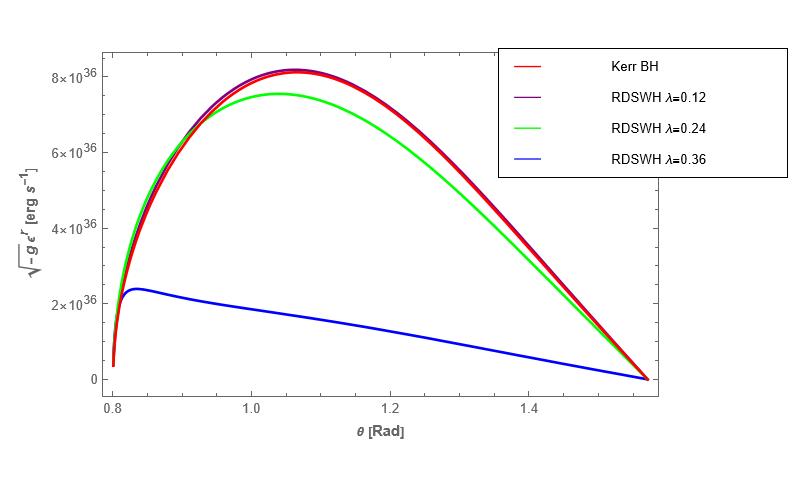}
    \caption{Electromagnetic energy flux per unit solid angle for spin parameter $a=0.97$, varying the deformation parameter.}
  \label{fig:a097}
\end{figure}

\begin{figure}
    \centering
    \includegraphics[width=0.9\textwidth] {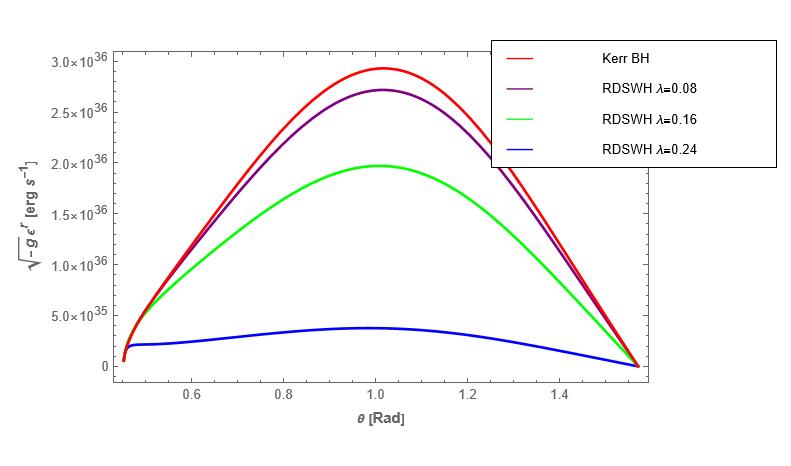}
    \caption{Electromagnetic energy flux per unit solid angle for spin parameter $a=0.99$, varying the deformation parameter.}
  \label{fig:a099}
\end{figure}

The electromagnetic energy extraction rate is calculated by integrating the flux through the wormhole $r_{\mathrm{ISCO}}$ into the solid angle
\begin{eqnarray}
    P_{\mathrm{BZ}}&=&\frac{2}{c^{3}}\int_{0}^{2\pi}d\phi\int_{\theta_{\mathrm{inicial}}}^{\frac{\pi}{2}}d\theta\sqrt{-g}\mathcal{E}^{r}(r_{\mathrm{ISCO}},\theta) \nonumber\\
    &=&\frac{4\pi}{c^{3}}\int_{\theta_{\mathrm{initial}}}^{\frac{\pi}{2}}d\theta\sqrt{-g}\mathcal{E}^{r}(r_{\mathrm{ISCO}},\theta).
    \label{4.28}
\end{eqnarray}

where $\theta_{{\mathrm{initial}}}$ is the value of the axial angle at which the flow $\mathcal{E}^{r}$ cancels out. This angle is determined only by the azimuthal component of the magnetic field (for a smaller value of the angle, $B^{\phi}$ is imaginary), and therefore it depends on the model of the magnetic field geometry that we have adopted\footnote{The angle  $\theta_{\mathrm{initial}}$ depends on the spin $a$. Its dependence with the deformation parameter $\lambda$ is minimal, as can be see in Figures \ref{fig:a095},\ref{fig:a097} and \ref{fig:a099}.}

We consider a wormhole with parameters similar to those of a stellar-mass black hole, with a mass $M=10\, M_{\odot}$ and a magnetic field $B_{0}=10^{7}$ G. We choose three different spin values in the range we established earlier, $a=$0.95 $GM/c^{2}$, $a=$0.97 $GM/c^{2}$ and $a=$0. 99 $GM/c^{2}$; for each of these cases we take four different deformation parameters: the minimum value is $\lambda=$0, which corresponds to a Kerr black hole, and the maximum value of the deformation parameter is such that the throat and the last stable circular orbit are close.

We plot in Figures \ref{fig:a095}, \ref{fig:a097} and \ref{fig:a099}  the integrand of expression \eqref{4.28} for different values of the spin and deformation parameters as a function of $\theta$. The curves present two different behaviors: first, as the spin parameter $a$ increases, the angular flux distribution becomes symmetric in the axial angle $\theta$; this unexpected behavior lies in the nonlinear dependence of the polar coordinate, present in the magnetic frequency $\omega(r,\theta)$ and the radial $B^{r}_{\mathrm{new}}(r,\theta)$ and azimuthal $B^{\phi}(r,\theta)$ components; and second, as the spin parameter $a$ decreases, the behavior of the curves changes, so that the Kerr case ceases to dominate in the face of slightly deformed wormholes. Furthermore, it can be seen that regardless of the value of the spin parameter $a$, the electromagnetic energy extraction rate drops significantly for highly deformed wormholes.

In the previous section, we defined the parameter $\hat{\lambda}(a)$ such that the innermost stable circular orbit coincides with the throat of the wormhole. In addition, the assumed magnetic field model is such that the radial component cancels out over the throat. Therefore, when calculating the flux \eqref{52} (which is proportional to this component of the magnetic field), given this choice of deformation parameter, the Poynting flux is expected to be zero. This is the main reason for the extreme drop observed for values of the deformation parameter sufficiently close to $\hat{\lambda}(a)$: since $r_{\mathrm{ISCO}}$ is quite close to $r_{+}$ (see Tables \eqref{table:er095}, \eqref{table:er097}, and \eqref{table:er099}), the radial component of the magnetic field at $r_{\mathrm{ISCO}}$ diminishes dramatically.

Notice from the tables that the maximum electromagnetic energy extraction rate $P_{\mathrm{BZ}}$ is obtained for $a = 0.97$ and $\lambda = 0.12$. On the other hand, as the spin value increases (in the analyzed range), the value of the maximum power moves to smaller values of the $\lambda$ parameter.

Finally, it should be noted that the results obtained here depend on the model adopted for the magnetic field geometry. Although the electromagnetic flux depends (roughly) on the square of the magnetic field modulus, the latter has a highly nonlinear dependence on the spin and the $\lambda$ parameter. Therefore, it is not possible to conclude in a general way whether the Blandford-Znajek mechanism is more efficient in a Kerr BH or in a DSRW; this can only be determined by a case-by-case analysis, fixing the parameters that fully determine both spacetimes: mass $M$ and spin $a$, and for the DSRW the deformation parameter $\lambda$.

\begin{table}
\large
\centering
\begin{tabular}{c | c | c} 
 $a=0.95 \;GM/c^2$ & $r_{+}/(GM/c^2$) & $P_{\mathrm{BZ}}$ [$10^{36}$ $erg\; s^{-1}$] \\ [0.5ex] 
 \hline\hline
 $\lambda$=0 & 1.31225 & 1.049  \\ 
 \hline
 $\lambda$=0.18 & 1.43657 & 1.172  \\
 \hline
 $\lambda$=0.35 & 1.72042 & 1.303  \\
 \hline
 $\lambda$=0.445 & 1.92793 & 0.300  \\
 \hline
\end{tabular}
\caption{Values of the energy extraction rates with mass $M=10\, M_{\odot}$, spin parameter $a$=0.95 $GM/c^2$, ISCO radius $r_{\mathrm{ISCO}}/(GM/c^2)=$1.93724 and magnetic field $B=10^{7}$ $G$, at different values of the deformation parameter $\lambda$.}
\label{table:er095}
\end{table}

\begin{table}
\large
\centering
\begin{tabular}{c | c | c} 
$a=0.97 \;GM/c^2$ & $r_{+}/(GM/c^2$) & $P_{\mathrm{BZ}}$ [$10^{36}$ $erg\; s^{-1}$] \\ [0.5ex] 
 \hline\hline
 $\lambda$=0 & 1.2431 & 4.080  \\ 
 \hline
 $\lambda$=0.12 & 1.31123 & 4.134  \\
 \hline
 $\lambda$=0.24 & 1.47905 & 3.836  \\
 \hline
 $\lambda$=0.36 & 1.70847 & 0.969  \\
 \hline
\end{tabular}
\caption{Values of the energy extraction rates with mass $M=10\, M_{\odot}$, spin parameter $a$=0.97 $GM/c^2$, ISCO radius $r_{\mathrm{ISCO}}/(GM/c^2)=$1.73752 and magnetic field $B=10^{7}$ $G$, at different values of the deformation parameter $\lambda$.}
\label{table:er097}
\end{table}

\begin{table}
\large
\centering
\begin{tabular}{c | c | c} 
$a=0.99 \;GM/c^2$ & $r_{+}/(GM/c^2$) & $P_{\mathrm{BZ}}$ [$10^{36}$ $erg\; s^{-1}$] \\ [0.5ex] 
 \hline\hline
 $\lambda$=0 & 1.14107 & 2.038  \\ 
 \hline
 $\lambda$=0.08 & 1.18734 & 1.910  \\
 \hline
 $\lambda$=0.16 & 1.29347 & 1.426  \\
 \hline
 $\lambda$=0.24 & 1.42965 & 0.293  \\
 \hline
\end{tabular}
\caption{Values of the energy extraction rates with mass $M=10\, M_{\odot}$, spin parameter $a$=0.99 $GM/c^2$, ISCO radius $r_{\mathrm{ISCO}}/(GM/c^2)=$1.4545 and magnetic field $B=10^{7}$ $G$, at different values of the deformation parameter $\lambda$.}
\label{table:er099}
\end{table}

\section{Conclusions}

The motivation of this work was to analyze whether it is possible to extract rotational energy in rotating wormholes in the form of a Poynting flow, analogous to the Blandford-Znajek mechanism in black holes, and to quantify this process. This is done using the Damour-Solodhukin metric, which describes a Kerr-type rotating wormhole. We first study whether the NEC are violated in this spacetime; we find that two of them are violated globally, and the third is violated locally. 

We then argue for the feasibility of the Blanford-Znajek mechanism in a rotating spacetime with one ergoregion, and calculate the emitted electromagnetic flux, which depends on the geometry of the magnetic field surrounding the wormhole. To this end, the first requirement is to select those values of the spin and deformation parameter configurations such that the ergosphere is outside the throat.

As a second step, we analyze the possible ranges for the spin $a$ and deformation $\lambda$ parameters to characterize the Poynting flux. In our model, we require not only that the ergosphere is outside the throat, but also that some part of the accretion disk (which carries the charges that are the source of the magnetic field) penetrates the ergosphere; in other words, we impose that the radius of the innermost stable circular orbit should be smaller than the radius of the ergosphere at the equatorial plane. The values of the parameters that satisfy these constraints are $0.94281 \leq a < 1$ and $0 \leq \lambda \leq \Tilde{\lambda}_{\mathrm{crit}}(a) \leq \lambda_{\mathrm{crit}} =ac^{2}/{2GM}$. 

In addition, we propose a particular model for the geometry of the magnetic field: away from the ergosurface, there is a uniform magnetic field model pointing in the polar direction. Within this region, the poloidal component of the magnetic field decreases linearly as we approach the wormhole throat, with a toroidal component developing simultaneously, so that the modulus of the magnetic field is conserved within the ergoregion.

Our main conclusion is that there is a Poynting flow associated with the accretion process of matter and fields in rotating wormholes; this is an original result that, to our knowledge, has not been previously reported in the literature. Importantly, our results depend significantly on the geometry of the proposed magnetic field, which is evident from the behavior of the energy extraction rates with the spin parameter. Due to the nonlinear dependence of the magnetic field on the spin and deformation parameters, we do not obtain a general behavior describing the relationship between the Poynting fluxes emitted by a Kerr black hole and a rotating DS wormhole. However, we observe that the Poynting flux is maximum  for spin values close to $a \approx 0.97$, and that the Poynting flux of a DS wormhole is comparable to that of a Kerr black hole.

Although we are working with a simple model for the geometry of the magnetic field, it is in principle possible to find solutions of the stream equation in the metric of a rotating Damour-Solodukhin wormhole. This would allow a detailed study of the behavior of the magnetic field, especially near the throat. This kind of research has not yet been done in rotating wormhole spacetimes.

Finally, this analysis could be extended to other rotating wormhole spacetimes, such as the Teo wormhole and the rotating Ellis wormhole, allowing for the comparison of electromagnetic fluxes between different wormhole models.

\section{Appendix}\label{appendix}

From the symmetries of a stationary, axially symmetric spacetime (which is asymptotically flat), and from certain properties imposed on the electromagnetic tensor $F$ \cite{1974Wald}, it can be shown that it has the form

\begin{equation}\label{fe}
    F=\frac{1}{2}B_{0}\left(d\psi+2\frac{a}{c}d\eta\right),
\end{equation}

where $\psi$ is the axial killing vector and $\eta$ is the temporal (single form) killing vector, and the letter d denotes the exterior derivative. These contravariant killing vectors are written as

\begin{eqnarray}
    \psi & = & \psi^{\mu}\partial_{\mu}= \left(0,0,0,1\right),\\
    \eta &= & \eta^{\mu}\partial_{\mu}=\left(1,0,0,0\right),
\end{eqnarray}

and their corresponding covariant expressions are

\begin{eqnarray}
    \psi_{\mu} & = & g_{\mu\nu}\psi^{\nu}=g_{\mu\phi}=\left(g_{t\phi},0,0,g_{\phi\phi}\right),\\
    \eta_{\mu} & = & g_{\mu\nu}\eta^{\nu}=g_{\mu t}=\left(g_{tt},0,0,g_{t\phi}\right).
\end{eqnarray}

Let $\varphi$ be a k-form,

\begin{equation}
    \varphi=gdx^{I}=gdx^{i_{1}}\wedge dx^{i_{2}}\wedge...\wedge dx^{i_{k}}.
\end{equation}

The exterior derivative of a k-form is defined as

\begin{equation}
    d\varphi=\frac{\partial g}{\partial x^{i}}dx^{i}\wedge dx^{I}.
\end{equation}

Then the electromagnetic tensor given by \eqref{fe} yields

\begin{eqnarray}
    F & = &\frac{1}{2}B_{0}\left[\frac{\partial}{\partial r}\left(g_{t\phi}+2\frac{a}{c}g_{tt}\right) dr\wedge dt+\frac{\partial}{\partial \theta}\left(g_{t\phi}+2\frac{a}{c}g_{tt}\right) d\theta\wedge dt \right. \nonumber \\
   & & + \left. \frac{\partial}{\partial r}\left(g_{\phi\phi}+2\frac{a}{c}g_{t\phi}\right)dr\wedge d\phi+\frac{\partial}{\partial \theta}\left(g_{\phi\phi}+2\frac{a}{c}g_{t\phi}\right)d\theta\wedge d\phi \right].
\end{eqnarray}

Denoting $X$ as

\begin{eqnarray}
    X&=&g_{\phi\phi}+2\frac{a}{c}g_{t\phi}=\frac{\sin^{2}{\theta}}{\Sigma}\left(\rho^{2}-4\frac{GM}{c^{2}}a^{2}r\right), \\
    \rho^{2}&=&(r^{2}+a^{2})^{2}-a^{2}\Delta\sin^{2}{\theta}, \\
    \Delta&=&r^{2}-\frac{2GM}{c^{2}}r+a^{2},
\end{eqnarray}

the radial and axial magnetic field components are

 \begin{eqnarray}
    B^{r}(r,\theta)=\frac{1}{\sqrt{-g}}F_{\theta\phi}=\frac{B_{0}}{2\Sigma\sin{\theta}}\sqrt{\frac{\hat{\Delta}}{\Delta}}\frac{\partial X}{\partial\theta}, \\
    B^{\theta}(r,\theta)=\frac{1}{\sqrt{-g}}F_{\phi r}=-\frac{B_{0}}{2\Sigma\sin{\theta}}\sqrt{\frac{\hat{\Delta}}{\Delta}}\frac{\partial X}{\partial r}.
\end{eqnarray}

\backmatter

\bmhead{Acknowledgments} 
D. P. acknowledges the support from CONICET under Grant No. PIP 0554 and AGENCIA I$+$D$+$i under Grant PICT-2021-I-INVI-00387.

\section*{Declarations}


\begin{itemize}
\item Funding
D. P. acknowledges the support from CONICET under Grant No. PIP 0554 and AGENCIA I+D+i under Grant PICT-2021-I-INVI- 00387.

\item Competing interests 

\item Ethics approval 

\item Consent to participate

\item Consent for publication

\item Availability of data and materials

\item Code availability

\item Authors' contributions All authors contributed equally to the manuscript.

\end{itemize}







\bibliography{sn-bibliography}

\end{document}